\documentclass[11pt]{article}
\pdfoutput=1
\usepackage{amsmath,amssymb,multirow,jheppub,graphicx}
\hypersetup{colorlinks,linkcolor=blue,citecolor=cyan}
\usepackage{graphicx}
\usepackage{amsmath,amssymb}
\usepackage{epstopdf}
\usepackage{cancel}
\usepackage{tikz}
\usetikzlibrary{decorations.pathreplacing,
                calligraphy}% had to be loaded after decorations.pathreplacing library
\tikzset{
B/.style = {decorate,
            decoration={calligraphic brace, amplitude=4pt,
            raise=5pt, mirror},% for mirroring of brace
            very thick,
            pen colour=black},
dot/.style = {circle, fill, inner sep=2pt, outer sep=0pt}
        }
\usepackage{comment}
\usepackage{color}%

\def\hhref#1{\href{http://arxiv.org/abs/#1}{arXiv:#1}} % in bibliography

\graphicspath{{./Figures/}}

\newcommand\qt{\widetilde{q}}

 \usepackage{stackrel,nicefrac}
 \usepackage{mathtools}

\allowdisplaybreaks

\def\tilde{\widetilde}

\newtheorem{myconj}[equation]{Conjecture}

\title
{The Chern-Simons Natural Boundary and Black Hole Entropy}
\author{Griffen Adams and Gerald V. Dunne}

\affiliation{Department  of Physics, University  of Connecticut, Storrs, CT, 06269-3046}

\emailAdd{griffen.adams@uconn.edu,gerald.dunne@uconn.edu} 

\abstract{
The method of resurgent continuation of transseries reveals a new correspondence between the $q$-series for enumerating degeneracies of quarter-BPS states in supersymmetric black holes and $\widehat{Z}$ invariants of Chern-Simons theory on a class of 3 dimensional orientation-reversed manifolds. 
}

 \date{\today}

\begin{document}

 \maketitle

\tableofcontents

\setcounter{tocdepth}{2}
%\tableofcontents

\section{Introduction}
Natural boundaries are dense closed barriers of singularities, and their existence in physical theories poses both computational and conceptual questions. Natural boundaries are known to occur in many physical systems, such as statistical physics \cite{baxter,andrews,Wu:1975mw,Orrick:2001zz}, superconformal field theory \cite{BMO95}, gauge theory \cite{CCKPS3d,Cheng:2018vpl,CDGG}, wall crossing phenomena \cite{Gaiotto:2009hg,Gaiotto:2010okc,DMZ12} and field theory scattering amplitudes \cite{Mizera:2022dko,Caron-Huot:2023ikn}. It has recently been proposed \cite{CDGG,ACDGO,ACDGO2} that resurgence leads to a new form of unique continuation, beyond analytic continuation, applied to resurgent transseries rather than to analytic functions. This approach takes advantage of the rigidity of transseries under the operations of analysis \cite{ecalle,Sauzin06,costin-book}, a feature referred to as "preservation of relations".
This idea of {\it resurgent continuation} has been successfully applied to crossing the natural boundary in Chern-Simons (CS) theory \cite{CDGG,ACDGO,ACDGO2}. 
CS theory is a topological quantum field theory defined on a 3 dimensional manifold $\mathcal M_3$, 
with topological invariants $\widehat{Z}(\mathcal M_3;q)$ that are naturally expressed in terms of $q$-series, where $q=e^{-\hbar}$ with $\hbar$ being the CS coupling parameter. The $\widehat{Z}(\mathcal M_3;q)$ invariants provide a non-perturbative completion of Chern-Simons theory \cite{GMP,Gukov:2016gkn,Gukov:2017kmk}. In known cases, the $\widehat{Z}(\mathcal M_3;q)$ invariants exhibit a natural boundary on the unit circle $|q|=1$.  For certain  manifolds $\mathcal M_3$, the $\widehat{Z}(\mathcal M_3;q)$ invariants are related to Ramanujan's mock theta functions, providing a direct link between quantum field theory and number theory, specifically the theory of mock modular forms \cite{watson,GM12,lawrence,Hikami04,Hikami04a,Andersen:2018khh,Cheng:2018vpl,folsom,Zag09,Zwe08}.
%In known cases, the $\widehat{Z}(\mathcal M_3;q)$ invariants exhibit a natural boundary on the unit circle $|q|=1$. 
This is a rich subject, revealing deep connections between quantum field theory, topology, number theory and representation theory \cite{Cheng:2018vpl,CCKPS3d,Cheng:2022rqr}.

The motivation for studying the crossing of the natural boundary in quantum field theory (QFT) is to understand more deeply the subtleties of the QFT Feynman path integral, especially for gauge theories \cite{Witten:2010zr}. For CS theory, the natural boundary question is of particular interest because under orientation reversal of the 3 dimensional manifold, $\mathcal M_3\to {\overline{\mathcal M_3}}$, the CS action changes sign. Therefore, we would physically expect that a well-defined CS path integral should satisfy the following condition:
\begin{eqnarray}
    \int \, \mathcal{D}A \, \exp\left[-\frac{1}{\hbar}\int_{\mathcal M_3} \mathcal{L}_{CS}(A)\right] \,\,\, \stackrel{?}{=} \,\,\, \int \, \mathcal{D}A \, \exp\left[\frac{1}{\hbar}\int_{\overline{\mathcal M_3}} \mathcal{L}_{CS}(A)\right]
    \label{eq:path-integral}
\end{eqnarray}
Since the $\widehat{Z}(\mathcal M_3;q)$ invariants are expressed in terms of the path integral, we would therefore expect that the $\widehat{Z}$ invariants of $\mathcal M_3$ and its orientation reversal, $\overline{\mathcal M_3}$, are related by:
\begin{eqnarray}
    \widehat{Z}(\mathcal M_3;q^{-1}) = \widehat{Z}\left(\overline{\mathcal M_3};q\right)
\end{eqnarray}
where $\widehat{Z}(\mathcal{M}_3;q^{-1})$ is an appropriate extension of $\widehat{Z}(\mathcal{M}_3;q)$ beyond the unit circle. This is an extremely strong boundary crossing condition that relates the interior and exterior of the unit circle $|q|=1$, and poses a nontrivial test for the Chern-Simons path integral.

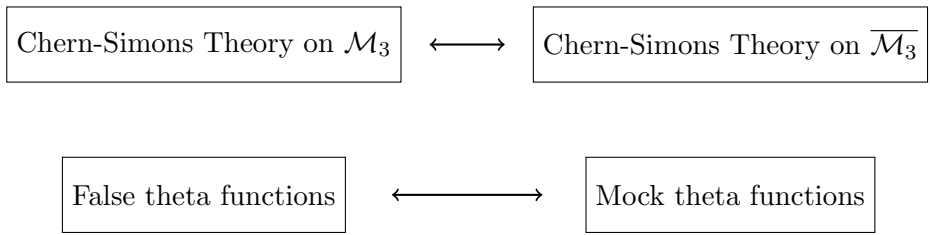
\begin{figure}[h]
\begin{tikzpicture}
  \draw (0,0) node {};
  \draw (4,-1) node[minimum height=1cm,minimum width=3.5cm,draw] {Chern-Simons Theory on $\mathcal{M}_3$};
  \draw[<->,thick] (7,-1) -- (8,-1);
  \draw (11,-1) node[minimum height=1cm,minimum width=3.5cm,draw] {Chern-Simons Theory on $\overline{\mathcal{M}_3}$};
  % \draw (4,-3) node[minimum height=1cm,minimum width=3.5cm,draw] {False theta functions};
  % \draw[<->,thick] (6.5,-3) -- (8.5,-3);
  % \draw (11,-3) node[minimum height=1cm,minimum width=3.5cm,draw] {Mock theta functions};
  \draw (4,-3) node[minimum height=1cm,minimum width=3.5cm,draw] {False theta functions};
  \draw[<->,thick] (6.5,-3) -- (8.5,-3);
  \draw (11,-3) node[minimum height=1cm,minimum width=3.5cm,draw] {Mock theta functions};
\end{tikzpicture}
  \caption{Symbolic representation of the two sides of the natural boundary, related by $q\to 1/q$.}
  \label{fig:boundarysides}
\end{figure}

This idea has been explored for a variety of 3-manifolds \cite{GMP,CCKPS3d,Cheng:2018vpl,CDGG,ACDGO}. It has been found by explicit computation that when $\mathcal M_3$ is a low-order Seifert homology sphere, such as $\Sigma(2,3,5)$ or $\Sigma(2,3,7)$, the CS $\widehat{Z}(\mathcal M_3;q)$ invariants are expressed in terms of false theta functions (roughly speaking,  theta-like functions with many "gaps": the summation being restricted by certain number theoretic congruences) \cite{BN19}. Then the boundary crossing identifies the $\widehat{Z}(\overline{\mathcal M_3};q)$ invariants for CS theory on the orientation-reversed manifold $\overline{\mathcal M_3}$ with certain mock theta functions (roughly speaking, theta-like functions with integer coefficients satisfying certain growth conditions). The $\widehat{Z}(\overline{\mathcal M_3};q)$ invariants are of interest because the integer coefficients are degeneracies of BPS states in certain superconformal field theories related to CS theory by the $3d$-$3d$ correspondence \cite{CDGS3d3d,Cheng:2018vpl,GMP}. This situation is represented symbolically in Figure \ref{fig:boundarysides}.

\vspace{0.3cm}

These results are extremely encouraging, but also quite puzzling, because there is only a limited class of Ramanujan mock theta functions for which the boundary crossing can be implemented with closed-form $q$-series expressions.
On the other hand, the physical expectation is that the correspondence in \eqref{eq:path-integral} should be more generally applicable. This means that there is something potentially interesting to be learned here about a gauge theory path integral.

This was the initial motivation for the papers \cite{CDGG,ACDGO,ACDGO2}.
There are general families of 3-manifolds $\mathcal M_3$ on one side of the natural boundary (the "unary side"), for which the $\widehat{Z}(\mathcal M_3;q)$ invariants can be expressed as special linear combinations of false theta functions \cite{GMP,Cheng:2018vpl,Chung20,CCKPS3d,GM21}. However, there is no known general closed-form expression for how a given false theta function crosses the natural boundary. Explicit expressions are only known for a limited class of very special linear combinations of false theta functions.
A numerical approach to fill this gap was proposed in \cite{CDGG,ACDGO,ACDGO2}, based on the idea of resurgent continuation. In the cases where mock theta expressions are known \cite{GM12}, this numerical approach produces the known $q$-series. Furthermore, it was shown in \cite{ACDGO} that it can also be applied beyond the known cases; for example, for any individual false theta function of weight $1/2$. As an application of this result, one can cross the natural boundary for CS theory on the general class of three-fibered Seifert homology spheres, $\Sigma(p_1,p_2,p_3)$, with mutually coprime $p_j$, thereby extending the known examples of  $\Sigma(2,3,5)$ or $\Sigma(2,3,7)$ studied in \cite{GMP,Cheng:2018vpl,CCKPS3d}. The resulting dual $q$-series have integer-valued coefficients with growth rates that have the physically expected Cardy-like behavior \cite{GJ23,ACDGO2,ceffmod}. Furthermore, they agree with the mock modular coefficients of the special mock Jacobi forms $\mathcal{Q}_M$, where $M=p_1p_2p_3$, appearing in Appendix A.1 of \cite{DMZ12}.

In this paper we extend the resurgent continuation approach to a new class of 3-manifolds, the Seifert manifolds with 4 singular fibers, $\mathcal M_3={\Sigma(p_1,p_2,p_3,p_4)}$. These manifolds are defined for integer-valued $p_k$ $(k=1, \dots, 4)$, and have particularly interesting topological and number theoretic properties when the $p_k$ are prime. For these manifolds, the CS $\widehat{Z}$ invariant has been constructed as a specific linear combination of false theta functions of weight $1/2$ and of weight $3/2$ \cite{Chung20}. The resurgent approach extends this analysis in three main directions:
\begin{enumerate}
    \item We extend the weight $3/2$ contribution to Chung's $\widehat{Z}$ invariant to a {\it vector-valued} set of false theta functions, $\widehat{Z}^{[j]}$, which we propose to identify with defect $\widehat{Z}$ invariants, as in \cite{CCKPS3d}. The new index $j=1, \dots, D$, where $D=\frac{1}{8}\prod_{k=1}^4 (p_k-1)$.
    
    \item Resurgent continuation takes this vector-valued set of false theta functions across the natural boundary, producing a vector of mock theta functions, each  with integer-valued coefficients. We propose to identify these as the new $\widehat{Z}^{[j]}(\overline{\mathcal{M}_3}, q)$ invariants for CS theory on the orientation reversed manifold $\overline{M}_3=\overline{\Sigma(p_1,p_2,p_3, p_4)}$.
    
    \item When the Seifert parameters 
    $\{p_1,p_2,p_3, p_4\}$ are distinct primes, the resulting mock theta functions have extra arithmetic structure, and for the lowest example, $\overline{\Sigma(2,3,5,7)}$, for which $D=6$, we find that the 6-vector of dual $q$-series after crossing the boundary agree precisely with $q$-series found by Dabholkar, Murthy and Zagier (DMZ) in their pioneering analysis \cite{DMZ12} of mock modular symmetry in the counting of degeneracies of quarter-BPS states in supersymmetric black holes, with charges related to $\{p_1,p_2,p_3, p_4\}$. We stress that DMZ were studying a very different problem from orientation reversal in Chern-Simons theory, and also used completely different techniques. This new correspondence suggests a much deeper connection between these two physical systems.
\end{enumerate}
These are the main results of this paper: we construct the dual $q$-series associated with the $\widehat{Z}^{[j]}$ invariants, $j=1, \dots 6$, for CS theory on
$\overline{\mathcal M_3}=\overline{\Sigma(2,3,5,7)}$ and show that they match the six independent $q$-series coefficients of the special mock Jacobi form $\mathcal{Q}_{210}$ that arises in the counting of degeneracies of quarter-BPS states of a particular supersymmetric black hole. These $\mathcal{Q}_M$ are identified by DMZ as crucial building blocks in the theory of mock Jacobi forms. In the black hole context, there is an important distinction between $\mathcal Q_M$ when the index $M$ is a product of an {\it even} or an {\it odd} number of primes \cite{DMZ12}. When $M$ has an even number of prime factors the $q$-series are "optimal" (in a sense defined below) and contribute to the counting of `immortal' single-centered black holes, while when $M$ has an odd number of prime factors there is only a finite number of cases where the $q$-series are optimal.
We show how to use resurgent continuation to construct the vector-valued  $\widehat{Z}^{[j]}$ invariants for any four fibered Seifert manifold with prime $p_i$, and we conjecture that they are related to the degeneracy counting for other examples of SUSY black holes associated with $\mathcal{Q}_{p_1p_2p_3p_4}$. This is represented symbolically in Figure \ref{fig:newresult} below.
\begin{figure}[h]
\begin{tikzpicture}
  \draw (0,0) node {};
  \draw (-6.6,0) node[minimum height=1cm,minimum width=3.5cm,draw] {$\widehat{Z}({\Sigma(p_1,p_2,p_3,p_4)};q)$};
  \draw[<->,ultra thick] (-4.4,0) -- (-3.5,0);
  \draw (-1.3,0) node[minimum height=1cm,minimum width=3.5cm,draw] {$\widehat{Z}(\overline{\Sigma(p_1,p_2,p_3,p_4)};q)$};
  \draw[<->,dashed,ultra thick] (0.9,0) -- (1.9,0);
  \draw (4.4,0) node[minimum height=1cm,minimum width=3.5cm,draw] {Coefficients of $\mathcal{Q}_{p_1p_2p_3p_4}$};
\end{tikzpicture}
  \caption{This paper adds an additional link to the correspondence in the first row of Table \ref{fig:boundarysides}. The double sided arrow on the left indicates that the $\widehat{Z}^{[j]}$ invariants on either side are related under $q \to 1/q$. The dashed arrow on the right indicates a precise matching between the Chern-Simons $\widehat{Z}^{[j]}$ invariants and the $q$-series coefficients $h_{(M,\ell_j)}$ of the decomposition of $\mathcal{Q}_{M}$ in \eqref{eq:key} on the black hole side, where $M=p_1p_2p_3p_4$.
  }
  \label{fig:newresult}
\end{figure}
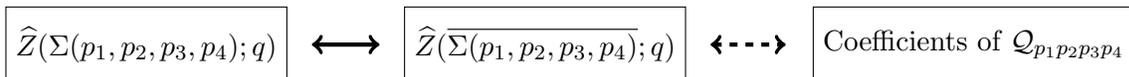

\section{BPS State counting in SUSY Black Holes}
\label{sec:bh}
We first recall the most relevant results in 
\cite{DMZ12,CD16} regarding the special mock Jacobi forms $\mathcal{Q}_M$ which appear as building blocks for the partition function of 
quarter-BPS dyons in an $\mathcal{N}=4$ supersymmetric $4d$ theory obtained through compactification of type II string theory on the product of a $K3$ surface and an elliptic curve \cite{DVV96}. The full partition function $\mathcal Z$ is the reciprocal of an object called the weight $10$ Igusa cusp form 
$\Phi_{10}(\tau, z, \sigma)$, which depends on three chemical potentials $(\sigma,\tau,z)$ that are conjugate to three $T$-duality invariant charges $(m,n,\ell)$. Here, $m$ and $n$ are associated with the magnetic and electric charge, respectively, and $\ell$ is a winding charge. The partition function has a Fourier decomposition,
\begin{eqnarray}
  \mathcal Z=  \frac{1}{\Phi_{10}(\tau, z, \sigma)} = \sum\limits_{m=-1}^{\infty}\psi_m(\tau,z)e^{2\pi i\, m\,\sigma}.
  \label{eq:partition}
\end{eqnarray}
With a suitable rescaling by Ramanujan's discriminant function\footnote{The discriminant function is defined in terms of the Dedekind eta function $\eta(\tau)$ as $\Delta(\tau)=\eta(\tau)^{24}$.}, $\Delta(\tau)$, the coefficients $\varphi_m(\tau,z)\sim\Delta(\tau)\psi_m(\tau,z)$ from the expansion \eqref{eq:partition} are weight $2$ Jacobi forms, meromorphic in $z$. A Jacobi form is a  function $\varphi(\tau,z):\mathbb{H} \times \mathbb{C} \to \mathbb{C}$ which transforms as \cite{DMZ12,folsom}
\begin{eqnarray}
    \varphi\left(\frac{a\tau+b}{c\tau+d},\frac{z}{c\tau+d}\right)&=&(c\tau+d)^k   e^{\frac{2\pi i m c z^2}{c\tau+d}}\varphi(\tau,z), \quad \begin{pmatrix} a & b \\ c & d \end{pmatrix} \in SL(2,\mathbb{Z}) \nonumber \\
    \varphi(\tau,z+\lambda \tau + \mu)&=& e^{-2\pi i m(\lambda^2 \tau +2\lambda z)} \varphi(\tau,z), \quad \lambda,\tau \in \mathbb{Z}
    \label{eq:jacobisymmetry}
\end{eqnarray}
where $m$ is the {\it level} of the Jacobi form, and $k$ is the {\it weight}. Modular symmetry is fundamental to these black hole problems \cite{Sen:1995in,Sen:2007qy,folsom}.  

In \cite{DMZ12} it is shown that there is a canonical decomposition of the $\varphi_m(\tau,z)$ into a polar term, $\varphi_m^P(\tau, z)$, which captures all of the poles in $z$, and term which is finite in $z$, $\varphi_m^F(\tau, z)$:
\begin{eqnarray}
    \varphi_m(\tau, z)=\varphi_m^F(\tau, z)+\varphi_m^P(\tau, z)
    \label{eq:decomp}
\end{eqnarray}
This decomposition has a physical interpretation for the black hole counting problem: the polar term has Fourier coefficients that count degeneracies of multi-centered black holes, while the coefficients in the finite term count immortal single-centered black holes. The polar piece $\varphi_m^P(\tau, z)$ has a known expression in terms of number-theoretic factors and an elementary Appell-Lerch sum (see equations (1.2)-(1.3) in \cite{DMZ12}). This Appell-Lerch sum exhibits the phenomenon of "wall-crossing": its Fourier coefficients, as an expansion in powers of $y=e^{2\pi i z}$, are $q$-series (where $q=e^{2\pi i \tau}$) whose coefficients jump when the chemical potentials cross walls of marginal stability in moduli space \cite{Gaiotto:2009hg,Gaiotto:2010okc,DMZ12,Alim:2010cf}. Physically, this describes the decay of the multi-centered black hole into {\it half}-BPS states as the walls in moduli space are crossed \cite{CV07}. Correspondingly, the modular symmetry of the finite part $\varphi_m^F(\tau,z)$ in \eqref{eq:decomp} is broken: they are {\it "mock"} modular, rather than truly modular.  {\it Mock modular} means that modularity is "weakly" broken, and can be restored by the addition of a non-holomorphic function known as the "shadow" \cite{DMZ12,Zag09,Zwe08}. This is an example of the "holomorphic anomaly", a competition between modularity and holomorphy of the generating function. The shadow provides a modular completion, but at the price of breaking holomorphy. For all of the examples in \cite{DMZ12}, the shadow is a modular form of weight $1/2$ or $3/2$. 

Due to the periodicity properties of mock Jacobi forms \eqref{eq:jacobisymmetry}, the finite part in \eqref{eq:decomp} has a further decomposition as a sum over theta functions 
\begin{eqnarray}
    \varphi_m^F(\tau,z) = \sum\limits_{\substack{\ell\in\mathbb{Z}/2m\mathbb{Z}\\(\ell,m)=1}} H_{(m,\ell)}(\tau)\, \vartheta_{m,\ell}(\tau,z)
    \label{eq:finite}
\end{eqnarray}
where the theta functions are defined as
\begin{eqnarray}
    \vartheta_{m,\ell}(\tau,z):= \sum\limits_{\substack{r \in \mathbb{Z} \\ r \equiv \ell \mod 2m}} q^{r^2/4m} y^r, \qquad q=e^{2\pi i\tau}, \,\, y=e^{2\pi i z}
\end{eqnarray}
The final step in this hierarchy of expansions in \eqref{eq:partition}-\eqref{eq:decomp}-\eqref{eq:finite} is that the coefficients $H_{(m, \ell)}(\tau)$ in \eqref{eq:finite} are mock modular forms with $q$-series expansions
\begin{eqnarray}
    H_{(m,\ell)}(\tau)=q^{-\ell^2/(4m)}\sum\limits_{n=0}^{\infty}A^{(m, \ell)}_n q^{n}
    \label{eq:mock-sum}
\end{eqnarray}
The physical significance of the expansion \eqref{eq:mock-sum} is that the integer-valued coefficients 
$A^{(m, \ell)}_n$ in \eqref{eq:mock-sum} enumerate degeneracies of quarter-BPS states in the $4d$ theory. There exists an ambiguity when forming the decomposition \eqref{eq:decomp}, as one can add any weak Jacobi form of the same weight and index to the polar piece without affecting its defining polar structure, which in turn changes the finite part $\varphi^F_m$. This ambiguity can be fixed to give a unique form for each weight and index by introducing an {\it optimality} condition on the $H_{(m,\ell)}$, which restricts the growth of the coefficients $A^{(m,\ell)}_n$ to be "as slow as possible" \cite{DMZ12,CD16}:
\begin{eqnarray}
    A^{(m, \ell)}_n \sim e^{\pi \sqrt{\Delta}/m} \qquad, \quad n\to\infty \qquad (\Delta:=4mn-\ell^2)
    \label{eq:optimalgrowth}
\end{eqnarray}
This optimality property is also important for the umbral moonshine program \cite{CDH12,CDH12a,Harrison:2022zee,Harvey:2019htf}, and has been further developed and refined in \cite{CD16}. Furthermore, we will see in the next section that this growth condition \eqref{eq:optimalgrowth} has a physical interpretation in the Chern-Simons context, and the factor in the exponential can be expressed in terms of the smallest Chern-Simons saddle action: see \eqref{eq:csgrowth}.

A key result of \cite{DMZ12} is the decomposition of the forms $\varphi^{\rm F}_m(\tau, z)$ into the special mock Jacobi forms $\mathcal{Q}_M(\tau, z)$, which have theta decomposition
\begin{eqnarray}
    \mathcal{Q}_M=\sum\limits_{\substack{\ell\in\mathbb{Z}/2M\mathbb{Z}\\(\ell,M)=1}} h_{(M,\ell)}(\tau) \vartheta_{M,\ell}(\tau,z), \qquad h_{(M,\ell)}(\tau) = q^{-\ell^2/(4m)}\sum\limits_{n=0}^{\infty} a_n^{(M,\ell)}q^n
    \label{eq:key}
\end{eqnarray}
The weight of the theta coefficients $h_{(M,\ell)}(\tau)$ of $\mathcal Q_M$ depends on the arithmetic properties of $M$. When $M$ has an {\it odd number} of prime factors, $\mathcal Q_M$  has mock modular coefficients $h_{(M,\ell)}(\tau)$ of weight $1/2$, which can be chosen to be optimal in only a finite number of cases, related to Ramanujan's mock theta functions. When $M$ has an {\it even number} of prime factors, $\mathcal Q_M$  has mock modular coefficients $h_{(M,\ell)}(\tau)$ of weight $3/2$. 
In this latter case, the corresponding $\mathcal{Q}_M(\tau, z)$ are special mock Jacobi forms of weight $2$, and enter into the decomposition of the finite part in \eqref{eq:decomp} as
\begin{eqnarray}
    \varphi_{m}^F(\tau,z) = \frac{1}{2^{\omega(m)-1}}\sum\limits_{\substack{M|m\\\mu(M)=+1}} \mathcal{Q}_M \big | \mathcal{V}^{(M)}_{2,m/M}
    \label{eq:phi2m}
\end{eqnarray}
Here $\omega(m)$ is the number of prime factors of $m$, and $\mu(M)=+1$ means that $M$ has an {\it even number} of distinct prime factors. The symbol $\mathcal{V}^m_{2,t}$ refers to a Hecke-like operator described in Section $4$ of \cite{DMZ12}, which mixes the coefficients of the mock modular forms $h_{(M,\ell)}$ according to the arithmetic properties of the integers $M$, $m$ and $t$.

By explicit computation \cite{DMZ12}, DMZ found that $\mathcal Q_1$ is optimal, and when $M$ is the product of two distinct primes they found that $\mathcal Q_6, \mathcal Q_{10}, \mathcal Q_{14}, \mathcal Q_{15}$ are also optimal in the sense that the discriminant bound, $\Delta:=4mn-\ell^2\geq -1$, is satisfied. This has the consequence that in the decomposition \eqref{eq:key} each theta-coefficient has the property that $h_{(M, \ell)}(\tau)\sim q^{-1/(4M)}$ as $q\to 0$. This implies that the coefficients grow as slowly as possible, consistent with the mock modular structure. Furthermore, DMZ found experimentally that this optimality property persisted for $\mathcal Q_M$ with $M$ being the product of two distinct primes up to $M=2*97=194$.

This motivated DMZ to consider the decomposition of $\varphi_{210}^F(\tau,z)$, where $m=2*3*5*7=210$ is the smallest product of {\it four} distinct primes, and where new structure arises which makes optimality less clear. For $m=210$, the decomposition \eqref{eq:phi2m} includes $\mathcal{Q}_{M}$ with $M=210$, and reads:
\begin{eqnarray}
    8\, \varphi^F_{210}(\tau,z)=\mathcal{Q}_1 \, | \, \mathcal{V}_{2,210}^{(1)} + \mathcal{Q}_6 \, | \, \mathcal{V}_{2,35}^{(6)} + \mathcal{Q}_{10} \, | \, \mathcal{V}_{2,21}^{(10)} + \mathcal{Q}_{14} \, | \, \mathcal{V}_{2,15}^{(14)} 
    \nonumber\\
    + \mathcal{Q}_{15} \, | \, \mathcal{V}_{2,14}^{(15)} + \mathcal{Q}_{21} \, | \, \mathcal{V}_{2,10}^{(21)} + \mathcal{Q}_{35} \, | \, \mathcal{V}_{2,6}^{(35)} + \mathcal{Q}_{210}
    \label{eq:phi210decomp}
\end{eqnarray}
Remarkably, after a "somewhat forbidding" computation involving finding the kernel of a very large matrix, DMZ found the surprising result that 
$\mathcal{Q}_{210}$ is also optimal. Specifically, they found 6 different Fourier coefficients $h_{(210,\ell_0)}(\tau)$, labeled by $\ell_0\in \left\{1, 11, 13, 17, 19, 23\right\}$, which are the first 6 numbers with ${\rm gcd}(\ell_0, 210)=1$. The other $\ell$ values up to 210 with ${\rm gcd}(\ell, 210)=1$ are related by symmetries to these six values of $\ell_0$. Table 4 in \cite{DMZ12} lists the coefficients of the $q$-series parts of the Fourier coefficients $h_{(210, \ell)}(\tau)$ for $\mathcal{Q}_{210}$, reproduced here for reference as Table \ref{tab:q210}.
  \begin{table}
  \centering
  \begin{tabular}{|p{0.5cm}|p{1cm}|p{0.5cm}|p{0.5cm}|p{0.5cm}|p{0.5cm}|p{0.5cm}|p{0.5cm}|p{0.5cm}|p{0.5cm}|p{0.5cm}|p{0.5cm}|p{0.6cm}|p{0.5cm}|p{0.5cm}|}
  \hline
  $\ell_0$ & $n=0$ & 1 & 2 & 3 & 4 & 5 & 6 & 7 & 8 & 9 & 10 & 11 &12\\ \hline
    1 & -2 & 13 & 28 & 34 & 49 & 46 & 71 & 59 & 83 & 77 & 102 & 87 & 121 \\
    11 & 0 & 9 & 21 & 27 & 36 & 41 & 51 & 44 & 75 & 62 & 62 & 82 & 104\\
    13 & 0 & 6 & 17 & 17 & 35 & 20 & 49 & 31 & 57 & 36 & 77 & 32 & 94\\
    17 & 0 & 4 & 12 & 16 & 22 & 19 & 43 & 17 & 40 & 50 & 41 & 27 & 87\\
    19 & 0 & 3 & 11 & 12 & 23 & 14 & 37 & 17 & 43 & 28 & 45 & 30 & 77 \\
    23 & 0 & 1 & 7 & 4 & 20 & -1 & 32 & 3 & 30 & 10 & 50 & -16 & 63\\ \hline
\end{tabular}
\caption{Table of the first 13 coefficients $a_n^{(M, \ell)}$ of the $q$-series part of the functions $h_{(M, \ell)}(\tau)$ in the theta decomposition of $\mathcal Q_M(\tau,z)$ in \eqref{eq:key}, for $M=2*3*5*7=210$. These coefficients are taken from 
Table 4 in \cite{DMZ12}.}
\label{tab:q210}
\end{table}
This means that the Fourier coefficients of the finite part $\varphi_m^F(\tau, z)$ actually combine into a vector-valued mock modular form (in this case of $M=210$, a vector of 6 mock modular forms), where each entry has coefficients satisfying the optimal growth property \eqref{eq:optimalgrowth}. 

In the next section we show that the very same $q$-series that arise in the theta decomposition of $\mathcal Q_M(\tau,z)$ in \eqref{eq:key}, for $M=2*3*5*7=210$ [as shown in Table \ref{tab:q210}], arise as $\widehat{Z}(\overline{\mathcal M_3}, q)$ invariants for Chern-Simons theory on the orientation reversal $\overline{\mathcal M_3}=\overline{\Sigma(2, 3, 5, 7)}$ of the Seifert manifold with four singular fibers $\mathcal M_3=\Sigma(2, 3, 5, 7)$. This suggests a deeper physical connection between the state counting problem for SUSY black holes and the topological invariants of Chern-Simons theory on orientation-reversed 3-manifolds. It also provides a new independent method of computing these $q$-series, as we describe in the next section. For completeness, we  also describe in Appendix \ref{sec:appendix1} how to use resurgent continuation to obtain the coefficients of $\mathcal{Q}_1|\mathcal{V}_{2,2p}^{(1)}$ for $p$ odd and $\mathcal{Q}_{p_1p_2}$ for $p_1$, $p_2$ being distinct primes. These account for all the other terms in \eqref{eq:phi2m}-\eqref{eq:phi210decomp} beyond $\mathcal Q_m$.

\section{Chern Simons $\widehat{Z}$ Invariants on 4-fibered Seifert homology spheres}
\label{sec:chernsimons}

In this section we extend the numerical resurgent continuation method proposed in \cite{ACDGO} to a new class of manifolds. We apply resurgent continuation to map the $\widehat{Z}(\Sigma(p_1,p_2,p_3,p_4);q)$ invariants to their orientation reversed counterparts. In the Chern-Simons context, $q=e^{-\hbar}$, where $\hbar$ is the CS coupling. A sketch of the procedure is:
\begin{enumerate}
    \item Identify the weight $3/2$ contribution to the $\widehat{Z}$ invariant of $\Sigma(p_1,p_2,p_3,p_4)$, as computed in \cite{Chung20}.
    \item Observe that the resulting combination of false theta functions can be expressed as the real part of the transseries decomposition of a particular combination of Mordell-Borel integrals, at the Stokes line $\hbar<0$. Compute the imaginary part of these integrals to identify \textit{other} false theta combinations, which we propose to identify with the defect $\widehat{Z}$ invariants, introduced in \cite{CCKPS3d}. This effectively extends $\widehat{Z}$ to a {\it vector} $\widehat{Z}^{[j]}$ of invariants in a way that captures the full effect of modular $SL(2,\mathbb Z)$ transformations. 
    \item 
    Express the transseries decomposition on the Stokes line as a unique decomposition into real and imaginary parts, and which is self-dual under the modular $S$ transformation.
    \item By the fundamental property of preservation of relations for resurgent transseries, impose the same transseries structure when rotated back across the natural boundary to $\hbar>0$. Then the numerical algorithm in \cite{ACDGO} can be applied to compute the integer coefficients of the dual $q$-series, which we identify as the weight $3/2$ contribution to the $\widehat{Z}^{[j]}$ invariants of the orientation reversed manifold $\overline{\Sigma(p_1,p_2,p_3,p_4)}$.
\end{enumerate}

The $\widehat{Z}$ invariants for Seifert manifolds with four singular fibers can be expressed as a linear combination of weight $1/2$ and weight $3/2$ false theta functions,
\begin{eqnarray}
\widehat{Z}\left(\Sigma(p_1,p_2,p_3,p_4);q\right)
&:=&\widehat{Z}_{3/2}(q)+\widehat{Z}_{1/2}(q), \qquad q=e^{-\hbar}
\\
&=&\sum\limits_{s=0}^{7}(-1)^s\left(\Phi^{(R_s)}_p(q)-R_s \Psi^{(R_s)}_p(q)\right)
\label{eq:zhat-split}
\end{eqnarray}
The $R_s$ in \eqref{eq:zhat-split} are integer-valued indices, defined below in \eqref{eq:Rindices}.
%that will be described later in the section. 
The building blocks in \eqref{eq:zhat-split} are the false theta functions $\Phi^{(a)}_p(q)$ and $\Psi^{(a)}_p(q)$ of weight $3/2$ and $1/2$, respectively:
\begin{eqnarray}    \Phi^{(a)}_p(q)&:=&\sum\limits_{n=0}^{\infty} n\, \varphi_{2p}^{(a)}(n)\, q^{n^2/4p}, \qquad \varphi_{2p}^{(a)}(n)=\begin{cases}
        1  \quad \text{ if }\,n\equiv \pm a  \mod{2p} \\ 0  \quad \text{ otherwise}
    \end{cases} 
    \label{eq:falsetheta32}\\
    \Psi^{(a)}_p(q)&:=&\sum\limits_{n=0}^{\infty} \psi^{(a)}_{2p}(n)\,q^{n^2/4p}, \qquad \psi_{2p}^{(a)}(n)=\begin{cases}
        \pm 1  \quad \text{ if }\,n\equiv \pm a  \mod{2p} \\ 0  \quad \,\,\,\,\text{ otherwise}
    \end{cases}
    \label{eq:falsetheta12}
\end{eqnarray}
The weight $3/2$ false theta functions will be the main focus of our work, and we can restrict to $1 \leq a \leq p-1$ by making use of the periodicity of these functions, $\Phi^{(a)}_p(q)=\Phi^{(-a)}_p(q)=\Phi^{(a+2p)}_p(q)$. The parameter 
\begin{eqnarray}
    p:=p_1 p_2 p_3 p_4
    \label{eq:p}
\end{eqnarray}
is the product of the 4 distinct prime numbers specifying the Seifert manifold.

The resurgent continuation method developed in \cite{ACDGO} allows one to construct the $\widehat{Z}$ invariants for \textit{orientation reversed} Seifert manifolds, $\overline{\Sigma(p_1,p_2,p_3,p_4)}$ given that the full $S$-transformation of the specific combination of false theta functions is known. In general, these combinations of false theta functions transform as vector-valued mock modular forms, and thus a key step in writing down the $S$-transformation is determining the specific vector of false theta functions. The methods in \cite{ACDGO} were developed for this purpose, identifying a natural analytic continuation of Mordell-Borel integrals whose transseries decompositions on the Stokes line produce $q$ and $\tilde{q}$ series that can be identified with false theta functions.

We start by defining a family of Mordell-Borel integrals
\begin{eqnarray}
    K_{(p,a)}(\hbar)&=&\frac{4}{\pi^2}\left(\frac{p \pi}{\hbar}\right)^{3/2}\int_0^{\infty} dx \, x \, e^{-p\, x^2/\hbar} \frac{\cosh{[(p-a) x]}}{\sinh{[p x]}}
    \label{eq:kpa}
    \\
    &=&\frac{p}{\pi^2}\int_0^{\infty} dx \, x \, e^{-p\,\hbar\, x^2/\pi^2} \frac{\sinh(2x)}{\cosh(2x)-\cos(a\pi/p)}
    \label{eq:kpa-dual}
\end{eqnarray}
which are similar to those in \cite{ACDGO} except with an extra factor of $x$ in the integrand, and the associated $\left(\frac{p \pi}{\hbar}\right)^{3/2}$ factor instead of $\left(\frac{p \pi}{\hbar}\right)^{1/2}$. Here, $p$ and $a$ are positive integers, with $1\leq a\leq p-1$.
These integrals have been defined for ${\rm Re}(\hbar)>0$, where the integrals are real. This is the region {\it inside} the natural boundary, where $|q|<1$. The dual expressions \eqref{eq:kpa-dual} are obtained from \eqref{eq:kpa} by Fourier transform \cite{CDGG,ACDGO}, using the basic identity
\begin{eqnarray}
  e^{-x^2/t}=\sqrt{\frac{t}{\pi}} \, \int_0^\infty dv \, e^{-v^2\, t/4}\, \cos(v\, x)
  \label{eq:fourier}
\end{eqnarray}
The small $x$ expansion of the trigonometric factors in the integrals in \eqref{eq:kpa} naturally generates the small $\hbar$ expansion of $K_{(p,a)}(\hbar)$, while the dual integrals in \eqref{eq:kpa-dual} naturally generate the large $\hbar$ expansion. 
Interestingly, {\bf both} the small and large $\hbar$ expansions are factorially divergent, which reflects the fact that the associated transseries representations of these integrals involve nonperturbative expansions in both $q=e^{-\hbar}$ and the modular dual $\tilde q=e^{-\pi^2/\hbar}$. See \eqref{eq:unarydecomposition} and \eqref{eq:nonunarydecomposition} below.

In analyzing the transseries structure, the richest source of information is on the Stokes line(s). Here the Stokes line
is the negative real axis $\hbar<0$ \cite{CDGG,ACDGO}, so to construct the transseries we rotate the integrals to the Stokes line. From the form of the integrals in 
\eqref{eq:kpa}-\eqref{eq:kpa-dual}, we see that the trigonometric Borel transforms are meromorphic in $x$, so under rotation to $\hbar<0$ the dual integrals in \eqref{eq:kpa-dual} produce a real series in powers of $q$, while the integrals in \eqref{eq:kpa} produce a purely imaginary series in powers of $\qt$, up to an overall rational power of $q$ and $\qt$, respectively.
We can approach the Stokes line from above or from below, defining the analytically continued integrals for $\hbar<0$ and $\epsilon \to 0$,
\begin{eqnarray}
    K^{(\pm)}_{(p,a)}(\hbar) = \mp\frac{4}{\pi^2}\left(\frac{p \pi}{\hbar}\right)^{3/2} i e^{\pm i\epsilon}\int_0^{\infty} dx \, (\mp i e^{\pm i \epsilon} x) \, e^{-p(\mp i e^{\pm i \epsilon} x)^2/\hbar} \frac{\cos{[(p-a) ( e^{\pm i \epsilon} x)]}}{\sin{[p ( e^{\pm i \epsilon} x)]}}
\end{eqnarray}
and similarly for the dual expression. The real and imaginary parts are found by averaging the rotated integrals above and below the Stokes line,
\begin{eqnarray}
    \text{Re}\left[K_{(p,a)}(\hbar<0)\right] &=& \frac{1}{2}\left(K^{(+)}_{(p,a)}(\hbar) - K^{(-)}_{(p,a)}(\hbar)\right) \\
    \text{Im}\left[K_{(p,a)}(\hbar<0)\right] &=& -\frac{i}{2}\left(K^{(+)}_{(p,a)}(\hbar) + K^{(-)}_{(p,a)}(\hbar)\right)
\end{eqnarray}
These reduce to straightforward residue calculations \cite{CDGG,ACDGO}. For even $p$ and $a$, the $q$-series coming from the real part is just the false theta function $\Phi^{(a)}_p\left(1/q\right)$, while the imaginary part will be a linear combination of certain other false theta functions of index $p$, as a function of the inverse  of the modular counterpart: $1/\tilde{q}$. 
\begin{figure}
    \centering
    \includegraphics[scale=.85]{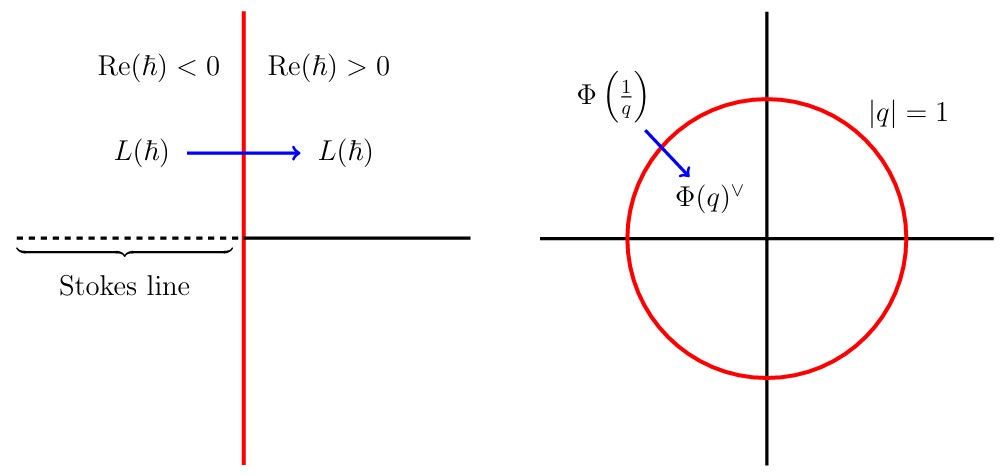}
    \caption{The $\hbar$ (left) and $q=e^{-\hbar}$ (right) complex planes. The natural boundaries of the false theta functions at $\text{Re}(\hbar)=0$, separating the left and right half planes, and $|q|=1$, bounding the unit circle, are denoted with red lines. The analytic continuation of the integrals from $\text{Re}(\hbar)<0$ to $\text{Re}(\hbar)>0$, which experiences no obstruction at the boundary, induces a continuation on the $q$-series as they cross over the unit circle from $|q|>1$ to $|q|<1$, leading to the \textit{dual} $q$-series functions $\Phi(q)^\vee$.}
    \label{fig:qhfigure}
\end{figure}
To begin this construction, we recognize that the linear combination of false theta functions appearing in the weight $3/2$ contribution to $\widehat{Z}(\Sigma(p_1,p_2,p_3,p_4);1/q)$
\begin{eqnarray}      \widehat{Z}_{3/2}\left(\Sigma(p_1,p_2,p_3,p_4);\frac{1}{q}\right) &=& \sum\limits_{s=0}^7 (-1)^s \Phi^{(R_s)}_p\left(\frac{1}{q}\right)
\end{eqnarray}
can be obtained as the real part of a linear combination of analytically continued Mordell-Borel integrals at $\hbar<0$. Note that we are considering the invariant as a function of $1/q$, since we are working in the domain $\hbar<0$. We also rescale the argument to $1/q^2$, which is reflected in the integrals by scaling $p$ and $a$ by a factor of two. This is just a matter of convenience, simplifying the analysis of the integrals and the subsequent numerics.
\begin{eqnarray}
    \sum\limits_{s=0}^7 (-1)^s \Phi^{(R_s)}_p\left(\frac{1}{q^2}\right)&=& \text{Re}\left[\sum\limits_{s=0}^7 (-1)^s K_{(2p,2|R_s|)}(\hbar)\right] 
    \label{eq:r1}\\
    &:=& \text{Re}\left[L_{(p,j^*)}(\hbar)\right]
    \label{eq:r2}
\end{eqnarray}
Here, the 8 possible values of $R_s$ are given by $p\left(\sum_{i=1}^4 \frac{\epsilon_i}{p_i}\right)$ with $\prod_{i=1}^4 \epsilon_i=(-1)^{s}$: see footnote $16$ in \cite{Chung20}. In order for the transseries decomposition of the integral to be self-dual under $\hbar\to\pi^2/\hbar$, which exchanges $q$ and $\tilde{q}$, we recognize that we must consider the analytic continuation of a \textit{vector} of Mordell-Borel integrals, where the real part of the entries forms a vector of false theta functions that contain exactly those appearing in the corresponding imaginary parts. This is the reason for the appearance of the index $j^*$ in $L_{(p,j^*)}(\hbar)$ - it denotes the distinguished entry in the vector of integrals (or false theta functions) that corresponds to the pure $\widehat{Z}$ invariant of the manifold, as constructed by Chung \cite{Chung20}.

The other $\widehat{Z}^{[j]}$ invariants, with $j\neq j^*$, are determined from the expressions on the Stokes line $\hbar<0$, by observing that the {\it imaginary part} is a sum over special combinations of false theta functions, but evaluated at $\tilde{q}$ instead of $q$:
\begin{eqnarray}
    \text{Im}\left[L_{(p,j^*)}(\hbar)\right] = \left(\frac{\pi}{\hbar}\right)^{3/2} \sum\limits_{j=1}^D M_{j^*j} \Phi^{[j]}_p\left(\frac{1}{\tilde{q}^2}\right), \quad \Phi^{[j]}_p(q) := \sum\limits_{s=0}^7 (-1)^s \Phi^{(R_s^{[j]})}_p(q)
    \label{eq:imag-l}
\end{eqnarray}
Here $D$ is the integer 
\begin{eqnarray}
    D=\frac{1}{8}\prod\limits_{i=1}^4 (p_i-1)
    \label{eq:d}
\end{eqnarray}
which encodes the length of the vector of false theta functions and integrals. The new false theta function combinations $\Phi^{[j]}_p$ in \eqref{eq:imag-l} depend on a new set of indices $R_s^{[j]}$, $j=1,\dots,D$, which generalize the $R_s$ indices in \eqref{eq:r1}-\eqref{eq:r2}. These can be used to construct the full $D$-component vector of integrals $L_{(p,j)}(\hbar)$ in a similar way as $L_{(p,j^*)}(\hbar)$. These integer-valued indices are given by 
\begin{eqnarray}
    R^{[j]}_s = 
        p\left(\sum_{i=1}^4 \frac{\ell_i\epsilon_i}{p_i}\right), \qquad 
            \prod_{i=1}^4 \epsilon_i=(-1)^{s} , \quad j=1,\dots,D
    \label{eq:Rindices}
\end{eqnarray}
and depend on a four dimensional lattice vector $\vec\ell^{[j]}=(\ell^{[j]}_1,\ell^{[j]}_2,\ell^{[j]}_3,\ell^{[j]}_4) \in \mathbb{Z}^4$. The construction of these for a given Seifert manifold is described in \cite{Hikami04}. These lattice vectors encode the upper indices of the false theta functions, the $R_s^{[j]}$ in \eqref{eq:imag-l}, and also the Chern Simons actions:
\begin{eqnarray}
{\rm Chern\,\, Simons\,\, actions:}\qquad    \Delta_j=r_j^2/4p
    \label{eq:csvalues}
\end{eqnarray}
Here $r_j:=\min_s\{|R_s^{[j]}|\}$ are the first $D$ integers coprime to all $p_i$. The Chern-Simons actions $\Delta_j$ are also the leading powers of the corresponding combination of false theta functions $\Phi_p^{[j]}(q)$.

For example, for the simplest four fibered Seifert homology sphere with distinct prime indices, $\Sigma(2,3,5,7)$, which by \eqref{eq:d} has $D=6$, we have the 6 lattice vectors:
\begin{eqnarray}
    \vec\ell^{[1]}&=&(1,1,2,3), \quad \vec\ell^{[2]}=(1,1,2,2), \quad 
    \vec\ell^{[3]}=(1,1,1,3)
    \nonumber\\
    \quad  \vec\ell^{[4]}&=&(1,1,1,2) ,\quad 
    \vec\ell^{[5]}=(1,1,2,1), \quad \vec\ell^{[6]}=(1,1,1,1)
\end{eqnarray}
These have been ordered according to increasing Chern-Simons actions:
\begin{eqnarray}
    \Delta_j \in \frac{1}{840}\{1^2,11^2,13^2,17^2,19^2,23^2\}
    \label{eq:deltas}
\end{eqnarray}
Observe that the $r_j\in \{1,11,13,17,19,23\}$ are the same integers that appear as $\ell_j$ in the decomposition of the special mock Jacobi form $\mathcal{Q}_{210}$; see the first column in Table \ref{tab:q210}. This is the first clear hint of a possible connection between the Chern-Simons boundary crossing problem and the black hole state-counting problem \cite{DMZ12}, described in the previous section.

The combinations of false theta functions in \eqref{eq:imag-l}, now evaluated at $q$, are the real parts of a vector of Mordell-Borel integrals when rotated to $\hbar<0$:
\begin{eqnarray}
    \text{Re} \left[L_{(p,j)}(\hbar)\right] = \Phi^{[j]}_p\left(\frac{1}{q^2}\right),\quad L_{(p,j)}(\hbar) := \sum\limits_{s=0}^7 (-1)^s K_{(2p,2|R^{[j]}_s|)}(\hbar), \quad j=1,\dots,D
\end{eqnarray}
These have similar imaginary decompositions to \eqref{eq:imag-l}. By computing the imaginary part of the {\it vector} comprised of the $L_{(p,j)}(\hbar)$ and grouping by the leading powers of $\Phi_p^{[j]}(1/\qt^2)$, one can extract the exact form of the mixing matrix $M_{jk}$, which in general will be built out of trigonometric factors $\sin(a\pi/p)$ for integer values of $a$ depending on $j$ and $k$.

Putting this all together, we have constructed the full transseries decomposition on the Stokes line of the vector of Mordell-Borel integrals $L_{(p,j)}(\hbar)$ in terms of the linear combination of false theta functions defined in \eqref{eq:imag-l}:
\begin{eqnarray}
    L_{(p,j)}(\hbar) = \Phi^{[j]}_p\left(\frac{1}{q^2}\right) + i\left(\frac{\pi}{\hbar}\right)^{3/2} \sum\limits_{k=1}^D M_{jk}\Phi^{[k]}_p\left(\frac{1}{\tilde{q}^2}\right)
    \label{eq:unarydecomposition}
\end{eqnarray}
This Stokes line decomposition is {\it unique}, because it is a transseries decomposition of the vector of integrals $L_{(p,j)}(\hbar)$ into real and imaginary parts. Notice that the real and imaginary parts involve the very same false theta combinations $\Phi^{[j]}_p$, but in terms of $q$ and its modular dual $\qt$, respectively. Consistency of the decomposition with the modular S transformation therefore implies that the
mixing matrix $M_{jk}$ is precisely the same mixing matrix in the behavior of these integrals under the modular S transformation $\hbar\to \pi^2/\hbar$:
\begin{eqnarray}
    L_{(p,j)}(\hbar) = \left(\frac{\pi}{\hbar}\right)^{3/2}\sum\limits_{k=1}^D M_{jk} L_{(p,k)}\left(\frac{\pi^2}{\hbar}\right)
    \label{eq:Lselfdual}
\end{eqnarray}
While these expressions were derived on the Stokes line $\hbar<0$, since this is the unique transseries decomposition, we therefore expect the same form of transseries decomposition on the other side of the natural boundary, where $\text{Re}(\hbar)>0$.  Here we invoke the "principle of preservation of relations" for transseries \cite{ACDGO}, based on the extreme structural rigidity of transseries \cite{ecalle,Sauzin06,costin-book}. Specifically, the mixing matrices must be the same, and the rational exponents of the $\Phi^{[j]}_p$ must also be preserved, even across a natural boundary.

Given that the $\hbar \to \pi^2/\hbar$ relation (\ref{eq:Lselfdual}) holds for all $\hbar \in \mathbb{C}$, the \textit{preservation of relations} of resurgent functions fixes many of the analytic properties of the decomposition in terms of the \textit{dual} $q$-series $\Phi^{[j]}_p(q)^\vee$. Crucially, the transseries decomposition into $q$ and $\tilde{q}$ series is preserved, with the dual $q$-series having the same overall rational power as their false theta counterparts,
\begin{eqnarray}
    \Phi^{[j]}_p(q)^\vee=q^{-\Delta_j}\sum\limits_{n=0}^\infty a^{(p,j)}_n  q^{n}, \qquad \Delta_j=\frac{r_j^2}{4p}.
    \label{eq:Phi-dual}
\end{eqnarray}
Physically, we expect the $a^{(p,j)}_n$ to be integers up to an overall normalization factor, given that they should encode the counting of BPS states. However, we do not enforce this in the numerical procedure: the $a^{(p,j)}_n$ turn out to be integers to very high precision, as for the weight $1/2$ analysis in \cite{ACDGO,ACDGO2}. 

In addition to preserving the rational exponents $\Delta_j$, we also must preserve the same mixing matrix and overall factor of $(\pi/\hbar)^{3/2}$ outside the $\tilde{q}$ terms. This motivates the following transseries decomposition of the vector of Mordell-Borel integrals $L_{(p,j)}(\hbar)$ also on the other side of the natural boundary, where $\hbar>0$:
\begin{eqnarray}
    L_{(p,j)}(\hbar) = \Phi^{[j]}_p\left(q^2\right)^\vee + \left(\frac{\pi}{\hbar}\right)^{3/2} \sum\limits_{k=1}^D M_{jk} \Phi^{[k]}_p\left(\tilde{q}^2\right)^\vee
    \qquad, \qquad \hbar>0
    \label{eq:nonunarydecomposition}
\end{eqnarray}
The left hand side of \eqref{eq:nonunarydecomposition} involves Mordell-Borel integrals that can be evaluated to extremely high precision. The right hand side involves the same $\Phi^\vee$, evaluated at $q^2$ and at $\tilde{q}^2$. Therefore, the coefficients $a_n^{(p, j)}$ in \eqref{eq:Phi-dual} can be fitted in the vicinity of the self-dual point $\hbar=\pi$, where $q=\tilde{q}$. This is a sensitive matching condition, which we use to solve numerically for the $a_n$ to high order in $n$, and with very high precision. This algorithm is described in detail in \cite{ACDGO}, where the procedure was implemented for Seifert manifolds with three singular fibers. 
The same fitting strategy is used here in this current paper. 

The transition from \eqref{eq:unarydecomposition} to \eqref{eq:nonunarydecomposition} implements resurgent continuation of the transseries structure from $\hbar\to -\hbar$, or $q \to 1/q$, effectively crossing the \textit{natural boundary} of the weight $3/2$ false theta functions. The output is a set of  natural candidates for the weight $3/2$ contribution to the $\widehat{Z}^{[j]}$ invariants of orientation reversed Seifert manifolds with four fibers.

For example, the resulting dual $q$-series associated to the orientation reversed four fibered Seifert manifold $\overline{\Sigma(2,3,5,7)}$ are:
\begin{eqnarray}
    \Phi^{[1]}_{210}(q)^\vee &=& 4\,q^{-1/840}\left(2 - 13 q - 28 q^2 - 34 q^3 - 49 q^4 - 46 q^{5} - 71 q^{6} - 59 q^{7}+\dots\right)\nonumber\\
    \Phi^{[2]}_{210}(q)^\vee &=& 4\,q^{-121/840}\left(-9 q - 21 q^2 - 27 q^3 - 36 q^4 - 41 q^{5} - 51 q^{6} - 44 q^{7}+\dots\right)\nonumber\\
    \Phi^{[3]}_{210}(q)^\vee &=& 4\,q^{-169/840}\left(-6 q - 17 q^2 - 17 q^3 - 35 q^4 - 20 q^{5} - 49 q^{6} - 31 q^{7}+\dots\right)
    \label{eq:2357}\\
    \Phi^{[4]}_{210}(q)^\vee &=& 4\,q^{-289/840}\left(-4 q - 12 q^2 - 16 q^3 - 22 q^4 - 19 q^{5} - 43 q^{6} - 17 q^{7}+\dots\right)\nonumber\\
    \Phi^{[5]}_{210}(q)^\vee &=& 4\,q^{-361/840}\left(-3 q - 11 q^2 - 12 q^3 - 23 q^4 - 14 q^{5} - 37 q^{6} - 17 q^{7}+\dots\right)\nonumber\\
    \Phi^{[6]}_{210}(q)^\vee &=& 4\,q^{-529/840}\left(- q - 7 q^2 - 4 q^3 - 20 q^4 + q^{5} - 32 q^{6} - 3 q^{7}+\dots\right)\nonumber
\end{eqnarray}
Notice first of all that the rational exponents in \eqref{eq:2357} 
are the Chern-Simons actions in \eqref{eq:deltas}, for which the $r_j$ values $\{1, 11, 13, 17, 19, 23\}$ coincide with the $\ell_0$ values in the first column of Table \ref{tab:q210} from the black hole state counting analysis of \cite{DMZ12}. 

Furthermore, comparing the $q$-series in \eqref{eq:2357} with the $q$-series coefficients in  Table $\ref{tab:q210}$, we see a striking relation between the $\Phi_{210}^{[j]}(q)^\vee$ and the coefficients $h_{(210,r_j)}(\tau)$ of the optimal mock Jacobi form $\mathcal Q_{210}$. In fact, we have
\begin{eqnarray}
    \Phi_{210}^{[j]}(q)^\vee = -4\, h_{(210,r_j)}(\tau), \quad q=e^{2\pi i \tau}, \, \tau \in \mathbb{H}
\end{eqnarray}
This gives an explicit relation between the weight $3/2$ contribution to the proposed $\widehat{Z}$ invariants of Chern-Simons theory on $\overline{\Sigma(2,3,5,7)}$ and the special mock Jacobi form $\mathcal Q_{210}$ used in the black hole microstate counting problem for $4d$ $\mathcal{N}=4$ theories \cite{DMZ12}, as reviewed in the previous section.

\begin{figure}[h!]
\centering
    \includegraphics[width=0.45\linewidth]{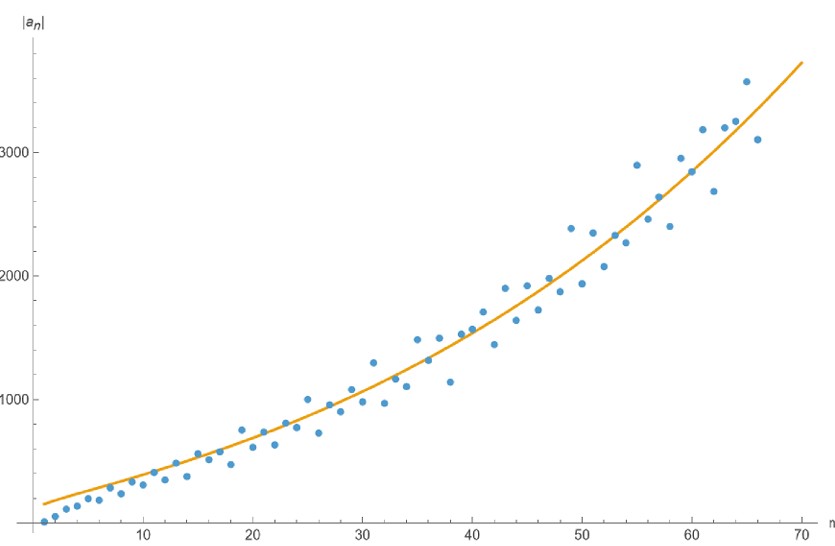}
    \includegraphics[width=0.45\linewidth]{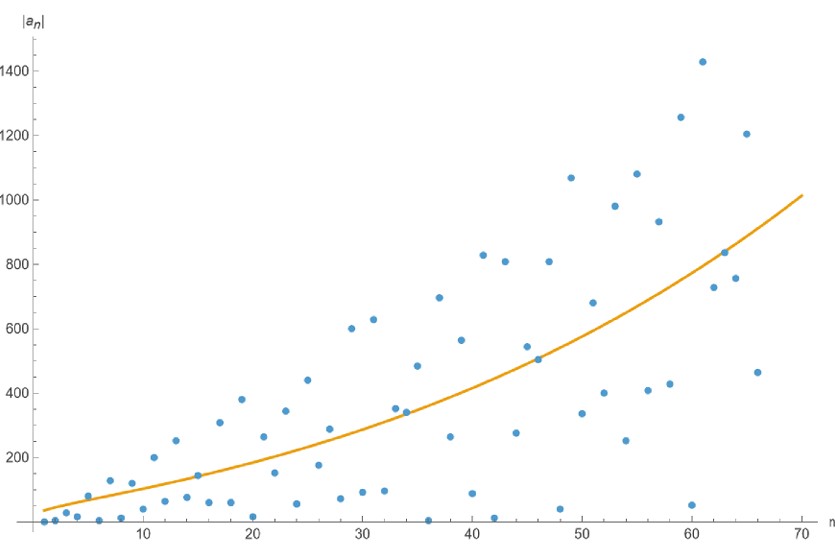}
\caption{Growth of the coefficients $a_n^{(210,j)}$ for $j=1$ (left) and $j=6$ (right). The larger spread of the coefficients in the $j=6$ case is due to the relative size of the subleading corrections, which increase as $j$ increases.}
\label{fig:phi210growth}
\end{figure}

In \cite{ACDGO2}, the growth of the coefficients of the dual $q$-series obtained by the resurgent boundary crossing method was analyzed for the weight $1/2$ case, providing a new approach to study the underlying 3d Cardy formula in the corresponding $\mathcal{N}=2$ $3d$ theory \cite{GJ23,ceffmod}. The approach of \cite{ACDGO2} extends straightforwardly to the weight $3/2$ $q$-series $\Phi_p^{[j]}(q)$:
\begin{eqnarray}
    a_n^{(p,j)} \sim e^{\pi \sqrt{16\Delta_1(n-\Delta_1)}}
    \qquad, \quad n\to\infty
    \label{eq:csgrowth}
\end{eqnarray}
where $\Delta_j=r_j^2/(4p)$, as in \eqref{eq:deltas}.
Our numerical results confirm this form of large order growth. Furthermore, this growth behavior matches the form of the large-order growth in the black hole analysis 
in  (\ref{eq:optimalgrowth}). See Figure \ref{fig:phi210growth}. Since these are "optimal", the growth is slow, but we also see that the scatter is greater for larger $j$, as expected from the analysis of subleading corrections, as described in Appendix C of \cite{ACDGO2}.

Our method can be applied to more general Seifert homology spheres with 4 singular fibers, $\Sigma(p_1, p_2, p_3, p_4)$, only limited by numerical precision considerations. For example, we record the results of our procedure for $p_k$ being 4 distinct primes, with the next smallest value of $p$ in \eqref{eq:p},  which is the Seifert  manifold $\Sigma(2,3,5,11)$. Here $p=330$ and $D=10$, and the Chern-Simons actions are:
\begin{eqnarray}
    \Delta_j \in \frac{1}{1320}\{1^2,7^2,13^2,17^2,19^2,23^2,29^2,31^2,47^2,49^2\}
\end{eqnarray}
Following the construction of the lattice vectors in \cite{Hikami04}, we find the following $10$ independent vectors giving the false theta indices \eqref{eq:Rindices}:
\begin{eqnarray}
    \vec\ell^{[1]}&=&(1, 1, 1, 4), \quad \vec\ell^{[2]}=(1, 1, 2, 5), \quad 
    \vec\ell^{[3]}=(1, 1, 2, 3), \quad  \vec\ell^{[4]}=(1, 1, 2, 2) ,\quad 
    \vec\ell^{[5]}=(1, 1, 1, 
  1) \nonumber \\
    \vec\ell^{[6]}&=&(1, 1, 2, 4), \quad \vec\ell^{[7]}=(1, 1, 1, 5), \quad 
    \vec\ell^{[8]}=(1, 1, 1, 3), \quad  \vec\ell^{[9]}=(1, 1, 2, 1) ,\quad 
    \vec\ell^{[10]}=(1, 1, 
  1, 2) \nonumber
\end{eqnarray}
This data allows us to construct the exact combination of Mordell-Borel integrals satisfying the $\hbar \to \pi^2/\hbar$ relation \eqref{eq:Lselfdual}. With this data we extract the transseries structure at the Stokes line and invoke preservation of this transseries structure back on the other side of the natural boundary. The expansion coefficients are then obtained from the non-unary transseries decomposition in \eqref{eq:nonunarydecomposition} by the same numerical fitting procedure in the vicinity of the self-dual point $q=e^{-\pi}$. We find the following dual $q$-series $\Phi^{[j]}_{330}(q)^\vee$ for $j=1, \dots, 10$:
\begin{eqnarray}
    \Phi^{[1]}_{330}(q)^\vee &=& \frac{4}{3}q^{-1/1320}\left(10-27 q-62 q^2-82 q^3-109 q^4-69 q^5-221 q^6-83 q^7+\dots\right)\\
    \Phi^{[2]}_{330}(q)^\vee &=& \frac{4}{3}q^{-49/1320}\left(-54 q - 119 q^2 - 123 q^3 - 203 q^4 - 177 q^5 - 236 q^6 - 234 q^7+\dots\right)\\
    \Phi^{[3]}_{330}(q)^\vee &=& \frac{4}{3}q^{-169/1320}\left(-41 q - 86 q^2 - 82 q^3 - 158 q^4 - 148 q^5 - 190 q^6 - 146 q^7+\dots\right)\\
    \Phi^{[4]}_{330}(q)^\vee &=& \frac{4}{3}q^{-289/1320}\left(-25 q - 49 q^2 - 85 q^3 - 94 q^4 - 89 q^5 - 128 q^6 - 139 q^7+\dots\right)\\
    \Phi^{[5]}_{330}(q)^\vee &=& \frac{4}{3}q^{-361/1320}\left(2 q - 28 q^2 - 6 q^3 - 59 q^4 + 28 q^5 - 137 q^6 + 54 q^7+\dots\right)\\
    \Phi^{[6]}_{330}(q)^\vee &=& \frac{4}{3}q^{-529/1320}\left(-40 q - 81 q^2 - 136 q^3 - 121 q^4 - 212 q^5 - 188 q^6 - 228 q^7+\dots\right)\\
    \Phi^{[7]}_{330}(q)^\vee &=& \frac{4}{3}q^{-841/1320}\left(-15 q - 58 q^2 - 63 q^3 - 136 q^4 - 46 q^5 - 187 q^6 - 142 q^7+\dots\right)\\
    \Phi^{[8]}_{330}(q)^\vee &=& \frac{4}{3}q^{-961/1320}\left(-7 q - 52 q^2 - 29 q^3 - 104 q^4 - 51 q^5 - 163 q^6 - 32 q^7+\dots\right)\\
    \Phi^{[9]}_{330}(q)^\vee &=& \frac{4}{3}q^{-2209/1320}\left(-9 q^2 - 14 q^3 - 68 q^4 - 3 q^5 - 71 q^6 - 56 q^7+\dots\right)\\
    \Phi^{[10]}_{330}(q)^\vee &=& \frac{4}{3}q^{-2401/1320}\left(-5 q^2 - 35 q^3 - 23 q^4 - 70 q^5 - 63 q^6 - 48 q^7+\dots\right)
\end{eqnarray}
We have computed up to $n=45$ coefficients and confirmed agreement with the leading asymptotic growth rate (\ref{eq:optimalgrowth}).
Motivated by these findings, we propose the following conjecture:
\begin{myconj}
    Let $\Phi_p^{[j]}(q)^\vee$ be the vector of mock modular forms dual under $q \to 1/q$ to the linear combination of false theta functions $\Phi^{[j]}_p(q)$, defined in \eqref{eq:imag-l}, associated to the $\widehat{Z}$ invariants of $\Sigma(p_1,p_2,p_3,p_4)$, where $p=p_1p_2p_3p_4$, and let $h_{(p,r_j)}(q)$ be the coefficients of the theta decomposition for the special mock Jacobi form $\mathcal Q_p$. Then, the following relation holds at the level of the $q$-series, 
    \begin{eqnarray}
        \Phi_p^{[j]}(q)^\vee = C \, h_{(p,r_j)}(q)
    \end{eqnarray}
    where $C \in {\mathbb Q}$ is a normalization constant.
\end{myconj}
To conclude this section, we make some final remarks and observations that are deserving of future study:
\begin{enumerate}
    \item In both of the examples we calculated, there is one $q$-series that does not follow the generic same-sign pattern of its coefficients as the other $q$-series in the vector. In both cases, this $q$-series is the dual of the original $\widehat{Z}$ invariant {\it without defects}, $\Phi^{[j^*]}_p(q)^\vee$. 
    \item By a corollary of Theorem $9.2$ in DMZ, the functions $\mathcal{Q}_M$, with $M=p_1p_2p_3p_4$, can always be chosen to have optimal growth. Thus, the dual $q$-series $\Phi^{[j]}_p(q)^\vee$, identified with the Fourier coefficients of $\mathcal{Q}_M$, will also have optimal growth for all $p=p_1p_2p_3p_4$, with coefficients growing as \eqref{eq:csgrowth}. By identifying the $\mathcal{Q}_M$ with the weight $3/2$ contribution to the $\widehat{Z}$ invariant, we can use the optimality property to make a general statement on the effective central charges of four-fibered Seifert homology spheres - see \cite{ACDGO2,ceffmod}.
    \item In \cite{Chung20}, expressions for Seifert homology spheres for $N$-fibers were found in terms of combinations of weights $1/2$ and $3/2$ false theta functions and their derivatives. Thus, in principle, using the methods of this paper and \cite{ACDGO}, resurgent continuation can be used to calculate the $\widehat{Z}$ invariant of Seifert homology spheres with $N$ fibers.
\end{enumerate}

\section{Conclusion}
In this work, we have extended the results of \cite{CDGG,ACDGO} for resurgent continuation across the Chern-Simons natural boundary to a new class of 3 dimensional manifolds: Seifert homology spheres with 4 singular fibers. These manifolds have a richer topological and arithmetic structure than their three-fibered counterparts, and have $\widehat{Z}$ invariants consisting of both weight 1/2 and 3/2 false theta functions. 
We extend Chung's $\widehat{Z}$ invariant \cite{Chung20} to a {\it vector} of  $\widehat{Z}^{[j]}$ invariants and derive its full structure under modular $SL(2, \mathbb Z)$ transformations. This is a similar strategy as was applied in \cite{ACDGO,ACDGO2} for weight 1/2 false theta functions for Seifert manifolds with 3 singular fibers. We focus here on the new weight 3/2 contribution, finding the full vector-valued false theta function structure on the Stokes line, expressed as a unique transseries decomposition of Mordell-Borel integrals into real and imaginary parts in terms of $q=e^{-\hbar}$ and its modular dual $\qt=e^{-\pi^2/\hbar}$. Invoking the principle of preservation of relations the resurgent continuation from the Stokes line across the natural boundary determines a vector of dual $q$-series which we identify with defect topological invariants for Chern-Simons theory on the orientation-reversed 4-fibered Seifert manifold.

In the lowest example with 4 distinct prime Seifert parameters, we find the surprising result that the resulting 6 new $q$-series on the orientation reversed manifold coincide precisely with the 6 $q$-series found in a different physical context: that of enumerating quarter-BPS states in supersymmetric black holes \cite{DMZ12}.  This reveals a new correspondence between the Chern-Simons topological invariants and the special mock Jacobi forms $\mathcal{Q}_p$, which are essential building blocks in the counting of quarter-BPS dyons in a $4d$ $\mathcal{N}=4$ theory obtained by string compactification. The boundary crossing method generalizes to Chern-Simons theory on other 4-fibered Seifert manifolds, and we conjecture that this provides a new way to compute degeneracies for these SUSY black holes. Future work will analyze further details of this web of dualities and correspondences.
\vskip 1cm 

\noindent{\bf Acknowledgement:} This material is based upon work supported by the U.S. Department of Energy, Office of Science, Office of High Energy Physics under Award Number DE-SC0010339. The authors thank O. Costin, S. Gukov and O. \"Oner for discussions and comments.

%\newpage 

\appendix

\section{Duals of weight $3/2$ false theta functions}
\label{sec:appendix1}
In this Appendix, we show that all of the mock Jacobi forms appearing in the decomposition \eqref{eq:phi210decomp} can be obtained as $q$-series dual to false theta functions under $q \to 1/q$. In particular, we show that the strongly holomorphic theta coefficients of $\mathcal{Q}_1|\mathcal{V}_{2,2p}^{(1)}$, i.e. those appearing in the decomposition of $\varphi^F_{2p}(\tau,z)$, can be obtained as duals of individual weight $3/2$ false theta functions, with a resurgent block structure similar to one studied in \cite{ACDGO} for false theta functions of weight $1/2$. Next, we show that the theta coefficients of $\mathcal{Q}_{p_1p_2}$, for $p_1$ and $p_2$ distinct primes,  can be obtained as the dual $q$-series of a combination of two weight $3/2$ false theta functions, with a procedure similar to that in Section \ref{sec:chernsimons} but with simpler arithmetic properties.

\subsection{Individual false theta functions duals}
We observed in Section \ref{sec:chernsimons} that the Mordell-Borel integrals $K_{(p,a)}(\hbar)$ in \eqref{eq:kpa} can be analytically continued to their Stokes line, $\hbar<0$, where they decompose into unique real and imaginary parts that can be written in terms of weight $3/2$ false theta functions $\Phi_{p}^{(a)}(q)$, defined in \eqref{eq:falsetheta32}. For odd $p$ (which we will restrict to for this section), and even $a$, their imaginary parts are sums over false theta functions in terms of $\qt$ with purely {\it even} upper indices, indicating that we can form a vector for a given $p$ which is self-dual under $\hbar \to \pi^2/\hbar$:
\begin{eqnarray}
    K_{(p,2j)}(\hbar)=\left(\frac{\pi}{\hbar}\right)^{3/2}\sum\limits_{k=0}^{\frac{p-1}{2}} M_{jk}\, K_{(p,2j)}\left(\frac{\pi^2}{\hbar}\right), \quad j=0,\dots,\frac{p-1}{2}
    \label{eq:Kselfdual}
\end{eqnarray}
While this structure makes it possible to apply resurgent continuation and solve for the dual $q$-series, it leaves out the false theta functions with odd upper index, $\Phi^{(2j+1)}_p(q)$. The imaginary parts of the corresponding Borel integrals $K_{(p,2j+1)}(\hbar)$ involve the $\Phi^{(2j+1)}_p$ evaluated at $\hbar \to \hbar + ip \pi$, acting on the integer powers of $q^{-a^2/4p}\Phi_{p}^{(a)}(q)$ as $q \to -q$. 
This structure is very similar to the {\it resurgent cyclic blocks} found in \cite{ACDGO}. Inspired by this, we introduce the class of Mordell-Borel integrals:
\begin{eqnarray}
    J_{(p,a)}(\hbar)= &=&\frac{2}{\pi^2}\left(\frac{p \pi}{\hbar}\right)^{3/2}\int_0^{\infty} dx \, x \, e^{-px^2/\hbar} \frac{\sinh{[(p-a) x]}}{\cosh{[p x]}}
    \label{eq:jpa}
\end{eqnarray}
Using these, we can define the vectors of integrals constituting the resurgent block:
\begin{eqnarray}
    L^{(A)}_{(p,j)}(\hbar) &=& K_{(p,2j)}(\hbar), \qquad j=0,...,\frac{p-1}{2} \\
    L^{(B)}_{(p,j)}(\hbar) &=& J_{(2p,2j+1)}(\hbar)+(-1)^{\frac{p-1}{2}+n} J_{(2p,2p-2j-1)}(\hbar), \qquad j=0,...,\frac{p-1}{2} \\ 
    L^{(C)}_{(p,j)}(\hbar) &=& K_{(p,2j+1)}(\hbar), \qquad j=0,...,\frac{p-1}{2} \\
    L^{(D)}_{(p,j)}(\hbar) &=& J_{(p,2j)}(\hbar), \qquad j=0,...,\frac{p-1}{2}
\end{eqnarray}
For all odd $p$, each of these vectors satisfies a modular relation under $\hbar \to \pi^2/\hbar$, similar to \eqref{eq:Kselfdual}. Sets $A$ and $B$ are self-dual, in the sense that they transform into themselves under $\hbar \to \pi^2/\hbar$, and sets $C$ and $D$ mix with each other:
\begin{eqnarray}
    L^{(A,B)}_{(p,j)}(\hbar)&=&\left(\frac{\pi}{\hbar}\right)^{3/2}\sum\limits_{k=0}^{\frac{p-1}{2}} M^{(A,B)}_{jk}\, L_{(p,k)}^{(A,B)}(\pi^2/\hbar) 
    \label{eq:LAB}\\
    L^{(C)}_{(p,j)}(\hbar)&=&\left(\frac{\pi}{\hbar}\right)^{3/2}\sum\limits_{k=0}^{\frac{p-1}{2}} M^{(C)}_{jk}\, L_{(p,k)}^{(D)}(\pi^2/\hbar)
    \label{eq:LCD}
\end{eqnarray}
where 
%$(\bullet)$ denotes the letters $A$, $B$, or $C$, and 
the mixing matrices $M^{(\bullet)}$, with $\bullet\in \{A, B, C\}$, depend on $p$ and have simple trigonometric entries. The transseries expansions found by a residue calculation on \eqref{eq:kpa}, \eqref{eq:jpa}, and their respective duals take the form
\begin{eqnarray}
    L^{(A)}_{(p,j)}(\hbar) &=& \epsilon_{j} \Phi^{(2j)}_p\left(\frac{1}{q}\right) + i \left(\frac{\pi}{\hbar}\right)^{3/2}\sum\limits_{k=0}^{\frac{p-1}{2}}M^{(A)}_{jk} \epsilon_{k}\Phi^{(2k)}_p\left(\frac{1}{\tilde{q}}\right), \quad \epsilon_j=\begin{cases}1, \quad j>0 \\ 2, \quad j=0\end{cases} 
    \label{eq:AB}\\
    L^{(B)}_{(p,j)}(\hbar) &=& \Phi^{(2j+1)}_p\left(\frac{i}{\sqrt{q}}\right) + i \left(\frac{\pi}{\hbar}\right)^{3/2}\sum\limits_{k=0}^{\frac{p-1}{2}}M^{(B)}_{jk} \Phi^{(2k)}_p\left(\frac{i}{\sqrt{\tilde{q}}}\right)
    \label{eq:BA}\\
    L^{(C)}_{(p,j)}(\hbar) &=& \Phi^{(2j+1)}_p\left(\frac{1}{q}\right) + i \left(\frac{\pi}{\hbar}\right)^{3/2}\sum\limits_{k=0}^{\frac{p-1}{2}}M^{(C)}_{jk} \Phi^{(2k)}_p\left(-\frac{1}{\tilde{q}}\right)
    \label{eq:LC}
\end{eqnarray}
Note that the matrices $M^{(\bullet)}$ in \eqref{eq:AB}-\eqref{eq:LC} are the same ones appearing in the $\hbar \to \pi^2/\hbar$ transformation law of the Mordell-Borel integrals in \eqref{eq:LAB}-\eqref{eq:LCD}. In practice these matrices are obtained through the residue calculation. To avoid notational clutter, the changes of variables in the argument of the false theta functions only acts on the integer power series obtained after factoring out the rational power of $q$. For example,
\begin{eqnarray}
    \Phi^{(2)}_3(q) = q^{1/12}(1 + 5 q^2 + 7 q^4 +\dots) \implies \Phi^{(2)}_3(iq) = q^{1/12}(1-5q^2+7q^4+\dots)
\end{eqnarray}
The same property of preservation of relations holds for these integrals and their transseries expressions, giving us a way to solve for the dual $q$-series $\Phi^{(a)}_p(q)^\vee$. For $\hbar>0$, the transseries expressions take the form
\begin{eqnarray}
    L^{(A)}_{(p,j)}(\hbar) &=& \epsilon_{j} \Phi^{(2j)}_p\left(q\right)^\vee + \left(\frac{\pi}{\hbar}\right)^{3/2}\sum\limits_{k=0}^{\frac{p-1}{2}}M^{(A)}_{jk} \epsilon_{k}\Phi^{(2k)}_p\left(\tilde{q}\right)^\vee, \quad \epsilon_j=\begin{cases}1, \quad j>0 \\ 2, \quad j=0\end{cases} \\
    L^{(B)}_{(p,j)}(\hbar) &=& \Phi^{(2j+1)}_p\left(i\sqrt{q}\right)^\vee +  \left(\frac{\pi}{\hbar}\right)^{3/2}\sum\limits_{k=0}^{\frac{p-1}{2}}M^{(B)}_{jk} \Phi^{(2k)}_p\left(i\sqrt{\tilde{q}}\right)^\vee \\
    L^{(C)}_{(p,j)}(\hbar) &=& \Phi^{(2j+1)}_p\left(q\right)^\vee + \left(\frac{\pi}{\hbar}\right)^{3/2}\sum\limits_{k=0}^{\frac{p-1}{2}}M^{(C)}_{jk} \Phi^{(2k)}_p\left(-\tilde{q}\right)^\vee
    \label{eq:firstblocknonunary}
\end{eqnarray}
It is interesting to note that the structure of this resugent block is \textit{identical} to that of the first resurgent block studied in \cite{ACDGO} for weight $1/2$ false theta functions, aside from the weight factor $(\pi/\hbar)^{3/2}$, instead of $(\pi/\hbar)^{1/2}$. 

When $p=1$, the integrals in sets $A$ and $C$ are, up to a normalization factor, identical to those appearing in the spinor and scalar diagonal heat kernel on $\mathbb{H}^2$ \cite{Dunne21}. The full set of transseries expressions \eqref{eq:firstblocknonunary} appears in Watson's work on the generating functions of class numbers. In fact, the dual $q$-series:
\begin{eqnarray}
    \Phi^{(0)}_1(q)^\vee &=& \frac{1}{4}\left(1-2q-4q^2-5q^3-\dots\right) \\
    \Phi^{(0)}_1(q)^\vee &=& -4q^{-1/4}\left(q^2+2q^4+3q^6+3q^8+\dots\right)
\end{eqnarray}
match, up to a constant factor, the generating functions of class numbers $\mathcal{U}(q)$ and $\mathcal{J}(q^2)$ in \cite{Watson35}.

For larger odd $p$, we find that the dual $q$-series $\Phi_p^{(a)}(q)^\vee$ are related to the mock Jacobi forms $\mathcal{Q}_1 | \mathcal{V}_{2,2p}^{(1)}$.
Let $\mathcal{Q}_1 | \mathcal{V}_{2,2p}^{(1)}$ have theta coefficients $h_{\ell}(q), \, \ell=0,...,2p$. For example, for $p=3$, we find that all the $h_\ell(q)$ from \cite{DMZ12} can be expressed in terms of the $\Phi_3^{(a)}$ as follows:
\begin{eqnarray}
    h_0(q) &=& -\frac{1}{2}(\Phi^{(0)}_3(q^{1/2})^\vee+\Phi^{(0)}_3(-q^{1/2})^\vee) \\
    h_1(q) &=& -\frac{1}{2} \Phi^{(1)}_3(-iq)^\vee \\
    h_2(q) &=& -\frac{1}{2}(\Phi^{(2)}_3(q^{1/2})+\Phi^{(2)}_3(-q^{1/2})) \\
    h_3(q) &=& -\frac{1}{2} \Phi^{(3)}_3(-iq)^\vee \\
    h_4(q) &=& -\frac{1}{2}q^{1/2}(\Phi^{(2)}_3(q^{1/2})-\Phi^{(2)}_3(-q^{1/2})) \\
    h_5(q) &=& q\, h_1(q) \\
    h_6(q) &=& -\frac{1}{2}q^{3/2}(\Phi^{(0)}_3(q^{1/2})-\Phi^{(0)}_3(-q^{1/2}))
\end{eqnarray}
Further studies show that this is a generic phenomenon: the theta coefficients of $\mathcal{Q}_1 | \mathcal{V}_{2,2p}^{(1)}$ can be identified in an alternating fashion with the dual $q$-series $\Phi^{(a)}_p(q)^\vee$ with even upper index and alternating combinations in terms of $q$ and $-q$ of $\Phi^{(a)}_p(q)^\vee$ with odd upper index. We have confirmed this for up to $p=13$, calculating roughly $n=40$ coefficients for each $q$-series and finding complete agreement with the $q$-series coefficients of $\mathcal{Q}_1 | \mathcal{V}_{2,2p}^{(1)}$, which can be computed by taking prescribed linear combinations of the coefficients of $\mathcal{Q}_1$, as described in Section 4 of \cite{DMZ12}.

\subsection{Combinations of two false theta functions duals}
Here we show how to find the theta coefficients of $\mathcal{Q}_{p_1p_2}$, appearing in the decomposition of $\varphi_{m}^F$, see \eqref{eq:phi2m}, when $m$ is the product of an even number of primes. These arise as the dual $q$-series of combinations of two false theta functions of weight $3/2$. The construction is far simpler than that described in Section \ref{sec:chernsimons}, and at index $p=p_1p_2$ requires the following vector of Mordell-Borel integrals:
\begin{eqnarray}
    N_{(p,j)}(\hbar) = \frac{1}{2}\left(K_{(2p_1p_2,2r_j)}(\hbar) - K_{(2p_1p_2,2\overline{r_j})}(\hbar)\right), \quad j=1,\dots,D
\end{eqnarray}
where the $K_{(p,a)}(\hbar)$ are defined in \eqref{eq:kpa}.
Here $r_j$ are the first $D=\frac{1}{2}(p_1-1)(p_2-1)$ integers coprime to $p_1$ and $p_2$, and $\overline{r_j}$ is the unique integer less than $p$ such that $r_j^2 \equiv \overline{r_j}^2 \mod{4p}$. These integrals are constructed precisely to obey a duality relation under $\hbar \to \pi^2/\hbar$:
\begin{eqnarray}
    N_{(p,j)}(\hbar)=\left(\frac{\pi}{\hbar}\right)^{3/2}M_{jk} \, N_{(p,j)}(\pi^2/\hbar), \quad j=1,\dots,D
    \label{eq:npjs}
\end{eqnarray}
where $M_{jk}$ is the mixing matrix, dependent on the indices $p$ and $r_j$. 
Continuing to the Stokes line $\hbar<0$, we find the unique $q$-series transseries decompositions into real and imaginary parts:
\begin{eqnarray}
    N_{(p,j)}(\hbar) = \overline{\Phi}^{[j]}_{p}\left(\frac{1}{q^2}\right) + i \left(\frac{\pi}{\hbar}\right)^{3/2} \sum\limits_{k=1}^{D} M_{jk} \overline\Phi_{p}^{[j]}\left(\frac{1}{\qt^2}\right)
    \label{eq:npj}
\end{eqnarray}
where we use the shorthand notation
\begin{eqnarray}
    \overline{\Phi}^{[j]}_{p}(q) = \Phi_{p}^{(r_j)}(q)-\Phi_{p}^{(\overline{r_j})}(q)
\end{eqnarray}
The matrix $M$ in \eqref{eq:npj} is the same that appears in the $\hbar \to \pi^2/\hbar$ relation \eqref{eq:npjs}, and in practice is constructed through the residue calculation. Since these integrals enjoy the analytic and resurgent properties as those described earlier, we can invoke the unique continuation under the property of preservation of relations, giving the transseries expressions for $\hbar>0$:
\begin{eqnarray}
    N_{(p,j)}(\hbar) = \overline{\Phi}^{[j]}_{p}\left(q^2\right)^\vee + \left(\frac{\pi}{\hbar}\right)^{3/2} \sum\limits_{k=1}^{D} M_{jk} \overline\Phi_{p}^{[j]}\left(\qt^2\right)^\vee
\end{eqnarray}
We illustrate this with the example of $(p_1,p_2)=(2,3)$, for which $D=1$. Implementing the numerical algorithm produces the $q$-series expansion
\begin{eqnarray}
    \overline{\Phi}^{[1]}_{6}(q)^\vee = \frac{1}{3}q^{-1/12}(1-35q-130q^2-273q^3-595q^4-1001q^5-\dots)
\end{eqnarray}
Up to an overall factor of $-4$, this matches the $q$-series expansion of the mock modular form $h_{(6,1)}(q)$, the (only) theta coefficient of the special mock Jacobi form $\mathcal{Q}_6$ in \cite{DMZ12}. We have tested that the generic relationship
\begin{eqnarray}
    \overline{\Phi}^{[j]}_{p_1p_2}(q)^\vee = -4 h_{(p_1p_2,r_j)}(q)
\end{eqnarray}
holds for all examples $\mathcal{Q}_{p_1p_2}$ in Appendix A.1 of DMZ \cite{DMZ12}, where $h_{(p_1,p_2,r_j)}(q)$ is the $j^{\text{th}}$ theta coefficient of the special mock Jacobi form $\mathcal{Q}_{p_1p_2}$. 

It is of interest to note that the integral $N_{(6,1)}(\hbar)$, which contains the mock modular coefficient $h_{(6,1)}$ of $\mathcal{Q}_6$ in its transseries expansion, can be reduced to the integral $T_{-1}^{(-)}(\hbar)$ found in Section 5.5 of \cite{CDGG}, which is the leading Borel integral in a double scaling limit that describes the small surgery limit of the $N=-1$ twist knot.

\end{document}